\documentclass[10pt]{article}

\usepackage{amsmath}
\usepackage{amssymb}
\usepackage{graphicx}

\usepackage{cite}

\usepackage{color}

\usepackage[labelfont=bf,labelsep=period,justification=raggedright]{caption}

\makeatletter
\renewcommand{\@biblabel}[1]{\quad#1.}
\makeatother

\date{}

\begin{document}

\begin{flushleft}
{\Large
\textbf{Gravity Effects on Information Filtering and Network Evolving}
}
\\
Jin-Hu Liu$^{1,2,3}$
Zi-Ke Zhang$^{2,3,\ast}$
Lingjiao Chen$^{1,2,3}$
Chuang Liu$^{2,3}$
Chengcheng Yang$^{1}$
Xueqi Wang$^{4,\ast}$
\\
\bf{1} Web Sciences Center, University of Electronic Science and Technology of China, Chengdu 610054,  People's Republic of China
\\
\bf{2} Institute of Information Economy, Hangzhou Normal University, Hangzhou, 311121, People's Republic of China
\\
\bf{3} Alibaba Research Center for Complexity Sciences, Hangzhou Normal University, Hangzhou, 311121, People's Republic of China
\\
\bf{4} Division of Translational Medicine, Shanghai Changzheng Hospital, Second Military Medical University, Shanghai, 200003, People's Republic of China
\\
$\ast$ E-mail: zhangzike@gmail.com, xueqiniu.wang@gmail.com
\end{flushleft}

\section*{Abstract}
In this paper, based on the gravity principle of classical physics, we propose a tunable gravity-based model, which considers
tag usage pattern to weigh both the mass and distance of network nodes. We then apply this model in solving the problems of information filtering and network evolving. Experimental results on two real-world data sets, \emph{Del.icio.us} and \emph{MovieLens}, show that it can not only enhance the algorithmic performance, but can also better characterize the properties of real networks. This work may shed some light on the in-depth understanding of the effect of gravity model.

\section*{Introduction}
As one of the four fundamental interactions of nature, the \emph{Gravity law} was discovered from the well-known Galilei's dropping ball experiment at the \emph{Leaning Tower of Pisa} \cite{Galileo1632}. In the next four centuries, it has proved a great success in explaining the basic mechanisms governing the revolution of the heavenly bodies in macroscopic space \cite{Newton1687}. In addition to understanding the natural rules, the \emph{Gravity model} has also been used in a wide rage of domains in discovering a universal mechanism driving the dynamics of various phenomena, such as population migration \cite{Karemera2000,Yuhongbo2008}, transportation flows \cite{Casey1955,Rietveld1989,De2001tm,Jung2008}, trade \cite{Tinbergen1962,Bergstrand1985,Rose2004,Westerlund2011} and trip~\cite{Reilly1929,Reilly1931,Celik2010} distributions. However, research on gravity-based interaction in online social systems is lacking attention. Recently, many pioneering works in this field focus on complex network based evolution and prediction, where nodes represent individuals, and edges denote the relations between them~\cite{Albert2002, Dorogovtsev2002, Newman2003, Boccaletti2006, Costa2007}. In network evolutionary studies, the main objective is to uncover the strategy of how to connect two nodes \cite{Albert2002}. There is a vast class of studies trying to understand both the static features and dynamic properties of evolving networks \cite{Newman2006, CuiAX2012}. Some classical models, such as \emph{ER} network \cite{ErdosP1960}, \emph{WS} network \cite{Watts199801}, \emph{BA} network \cite{Barabasi199901}, have been proposed to open the new horizons for the theoretical study of random graphs. After that, many extensive variants considering different factors (e.g. the aging effect \cite{Dorogovtsev2000, Dorogovtsev2007} and social impact \cite{Jin2001, Newman2003, Castellano2009}) have been presented to complete such a popular field. Analogously, the \emph{Information Filtering} technique aims at mining missing links via estimating the indirect similarities of the two observed nodes \cite{Clauset2008}. \emph{Recommender Systems} (\emph{RS})\cite{Lv2012} is one of the most promising information filtering techniques to solve the problem of \emph{Information Overload}. The \emph{RS} aims at finding objects (e.g. books, movies etc.) that are most likely to be collected by online users based on their historical behaviors and the attributes of nodes. Unlike the classical information retrieval strategy which can be viewed as recommending documents with given words \cite{Salton1983}, \emph{RS} can be classified into two categories: (i) estimation of similarity based on the historical records of user activities, such as user-based and object-based similarity \cite{Zhou2010,Liu2010,Liujg2011,LLY2011,Qiutian2011,Qiutian2013,QiuT2013alleviating, QiuT2014}; (ii) incorporating accessorial information, such as object attributes and descriptions, to extensively assist the corresponding prediction algorithms \cite{Zhangzk2010, Zhangzk20102}.

Therefore, the essential problem of both network evolution and recommender systems is to evaluate the similarity of each unconnected node pair, which is the core function that gravity model can provide. However, to the best of our knowledge, examples are relatively rare in adopting gravity model in online systems. In this paper, we apply gravity model in a particular scenario, \emph{Social Tagging Networks}, \cite{Zhangzk2010,Zhangzk20102, Zhangzk20101, Zhangzk20121,Zhangzk20122, Hufeng2013}, and take the tag usage pattern to weigh the nodes' mass and distance, and then verify this definition in a tunable classical gravity model. Experimental results on two representative datasets, \emph{Del.icio.us} and \emph{MovieLens}, show that the proposed gravity model can significantly enhance the recommendation performance. Further numerical observation on an evolutionary network model demonstrates that the gravity-based mechanism can better characterize the properties of real networks than the other two baseline models.

\section*{Gravity Based Recommender Systems}
We begin our study with introducing gravity law based recommendation algorithms, as well as two baseline algorithms to evaluate its performance on tag-based information filtering. Conventionally, a tag-aware recommender system can be represented in a triple form \cite{Zhangzk20102}: $G(U,O,T)$, where $U=\{U_1,U_2,\ldots,U_n\}$, $O=\{O_1,O_2,\ldots,O_q\}$, and $T=\{T_1,T_2,\ldots,T_r\}$ are respectively the sets of users, objects and tags. As a complete tagging action, $\mathbb{F}$, normally consists an arbitrary number of tags, e.g. $\mathbb{F}=\{U_1, O_1, T_1,T_2,T_3\}$, which indicates that user $U_1$ has assigned object $O_1$ with a tag set $\mathbb{T}=\{T_1,T_2,T_3\}$. Therefore, $\mathbb{T}$ can be regarded as attributes for both $U_1$ and $O_1$. Consequently, we use two matrices, $A^{uo}$ and $A^{ot}$, to describe the user-object and object-tag relations, respectively. For $A^{uo}$, if user $U_i$ has selected object $O_j$, $a_{ij}^{uo}=1$, otherwise $a_{ij}^{uo}=0$. Analogously, $a_{jk}^{ot}=1$, if $O_j$ has been assigned with tag $T_k$, and $a_{jk}^{ot}=0$, otherwise. In addition, we also use two weighted matrices, $B^{u}$ and $B^{o}$, to represent tagging preference of users and objects, respectively. We denote $b_{ik}^{u}$ as the number of tag $T_k$ assigned by user $U_i$, and $b_{jk}^{o}$ as the number of tag $T_k$ assigned with object $O_j$.

In this section, we introduce two baseline tag-aware algorithms based on the concept of mass diffusion \cite{Zhou2010}, as well as the proposed gravity law based algorithm. Given a target user $U_i$, the final resource of an object $O_j$, $f_{o_j}$, is calculated based on following methods. Finally, objects that $U_i$ hasn't selected will be recommended according to their respective final resources.

\par(I) Suppose the initial resource averagely located on objects $U_i$ has selected (resource for every selected object is initially set to 1) and each object equally distributes its resource to all neighbouring tags, and then each tag redistributes the received resource averagely to all its neighbouring objects. Therefore, after diffusion, the resource located on $O_j$ finally is \cite{Zhangzk2010}

\begin{equation}
f^{(I)}_{o_j}=\sum_{l=1}^r\sum_{s=1}^q\frac{a_{jl}^{ot}a_{sl}^{ot}a_{is}^{uo}}{k(T_l)k(O_s)}, j=1,2,\cdots,q,
\end{equation}
where $k(T_l)=\sum_{j=1}^qa_{jl}^{ot}$ is the number of neighboring objects of tag $T_l$, and $k(O_s)=\sum_{l=1}^ra_{sl}^{ot}$ is the number of neighbouring tags of object $O_s$.

\par(II) With considering the weighted matrix $B^{u}$, the initial resources are located on tags and the resource $T_j$ received is proportional to $b^{u}_{ij}$. Then each tag equally distributes the initial resource to all its neighbouring objects. Therefore, after diffusion, the resource located on $O_j$ finally is \cite{Zhangzk20102}

\begin{equation}
f_{o_j}^{(II)}=\sum_{l=1}^r\frac{a_{jl}^{ot}b_{il}^{u}}{k(T_l)}, j=1,2,\cdots,q,
\end{equation}
where $k(T_l)=\sum_{j=1}^qa_{jl}^{ot}$ is the number of neighboring objects of tag $T_l$.

\par(III) Different from (I) and (II), this algorithm dose not only consider the network structure, but also take into account the common features of both users and objects. In this paper, we adopt the gravity model to estimate the likelihood of each user-object pair. Based on the classical gravity model, the resource located on $O_j$ finally is

\begin{equation}
 f_{o_j}^{(III)}=\frac{m_{u_i}m_{o_j}}{(\frac{1}{C_{u_io_j}+\alpha})^2}, j=1,2,\cdots,q, \label{eq:GR}
\end{equation}
where $m_{u_i}=\sum\limits_{k=1}^{r}b_{ik}^{u}$ is the \emph{mass} of user $U_i$, $m_{o_j}=\sum\limits_{k=1}^{r}b_{jk}^{o}$ is the \emph{mass} of object $O_j$, $C_{u_io_j}=\sum\limits_{k=1}^{r}\delta_{ij}^k$, where $\delta_{ij}^k=1$ if $b_{ik}^{u}*b_{jk}^{o}\not=0$ and $\delta_{ij}^k=0$ otherwise, indicating the number of common properties that user $U_i$ and object $O_j$ both hold, and $\alpha \in [0,\infty)$ is a tunable parameter. $C_{u_io_j}$ only counts how many common tag attributes that user $U_i$ and object $O_j$ simultaneously have, neglecting the accumulated times each tag has been used neither by $U_i$ nor $O_j$. 

For one typical personalized recommendation process, it aims at optimizing the utility of each individual. That is to say, once the target user $U_i$ is fixed, the sole purpose of a recommendation algorithm is to estimate the score of every object, which is defined by on of the three objective functions Eq. (1) - Eq. (3). Therefore, the vector element of $f_j$ does not contain $U_i$ due to that the inial influence of $U_i$ is the same for every object. Once the object score function is defined, the recommendation process will be performed on each target individual $U_i$, and all the objects that s/he has not collected are ranked in a descending order according to $f_{o_j}$ (generated by any one of the three algorithms (I)-(III)). Eventually, the top $L$ objects will be recommended to this user.

Generally, algorithm (I) only considers the tags effect on objects, neglecting the users' interest in tags, algorithm (II) only considers the user-tag weights, neglecting the weights of object-tag relations, while algorithm (III) takes both of them into account. The advantages of algorithm (III) are clear. On one hand, algorithm (III) not only considers the popularity (reflected by mass) of users and objects, but also directly takes into account the common features between them. Hence it might be a promising way to mine the potential preference of users. On the other hand, algorithm (III) alternatively compares the tag attributes vector statistically, while both algorithms (I) and (II) focus on studying the diffusion process on tripartite networks . Therefore, it can clearly save computational cost by avoiding multi-step iterations.

\subsection*{Experimental Results}
\label{sec:experiment}
The empirical data we use in this paper include (datasets are free to download as \textbf{Supplementary Material}): (a) \emph{MovieLens}: one representative website, provided by GroupLens project (http://www.grouplens.org/), where users can vote movies in five discrete ratings 1-5; (b) \emph{Del.icio.us} (http://www.delicious.com/): obtained by downloading publicly available data from the social bookmarking website, which allows users to store, organize and retrieve personal bookmarks via social tags. To eliminate the data sparsity effect, in both datasets, we purify the data to guarantee that \cite{ZhangCX201301} (a) each user has collected at least one object; (b) each object has been collected by at least two users, and assigned by at least two tags; (c) each tag is used by at least two users. Table \ref{tab:data} shows the basic statistics of the observed data sets.

\par To test the algorithm performance, we randomly remove $10\%$ of the data as testing set and apply the algorithms in the remaining data to produce recommendations. In addition, to give a solid and comprehensive evaluation of the proposed algorithm, we employ three representative metrics to characterize the recommendations performance. (a) \emph{$AUC$} \cite{Hanely1982}, defined as the probability that the score of an examined link in the testing set is larger than those in the training set; (b) \emph{$Precision$} \cite{Sarwar2000}, defined as the successful ratio of the number of top $L$ recommended links divided by the recommendation length $L$; (c) \emph{Inter Similarity} \cite{ZhoutEPl2008}, defined as,  $InnerS=\frac{1}{n}{\sum_{i=1}^{n}}(\frac{2}{L(L-1)}\sum_{j\not=l}sim_{jl})$, where $sim_{jl}={\sum_{s=1}^{n}}(a_{sj}^{uo}a_{sl}^{uo})/\sqrt{\sum_{s=1}^n a_{sj}^{uo} \sum_{s=1}^n a_{sl}^{uo}}$ where $s$ runs over all users, is the similarity of object $O_i$ and $O_j$ appearing in the recommendation list for $U_l$.

\begin{figure}[!ht]
\begin{center}
\includegraphics[width=13cm]{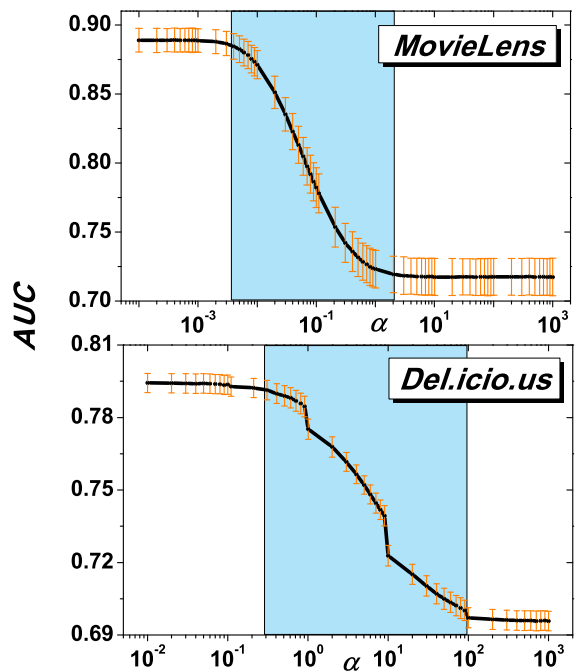}
\caption{(Color online) \emph{AUC} vs. $\alpha$ for algorithm (III) on the two observed datasets. The result is obtained by averaging over 50 independent realizations of random data division, and yellow lines represent the error intervals. It can be clearly seen that, for both datasets, \emph{AUC} decreases monotonously with $\alpha$, and reaches saturation for both large and small $\alpha$. }\label{Fig1}
\end{center}
\end{figure}

\begin{figure}[!ht]
\begin{center}
\includegraphics[width=12cm]{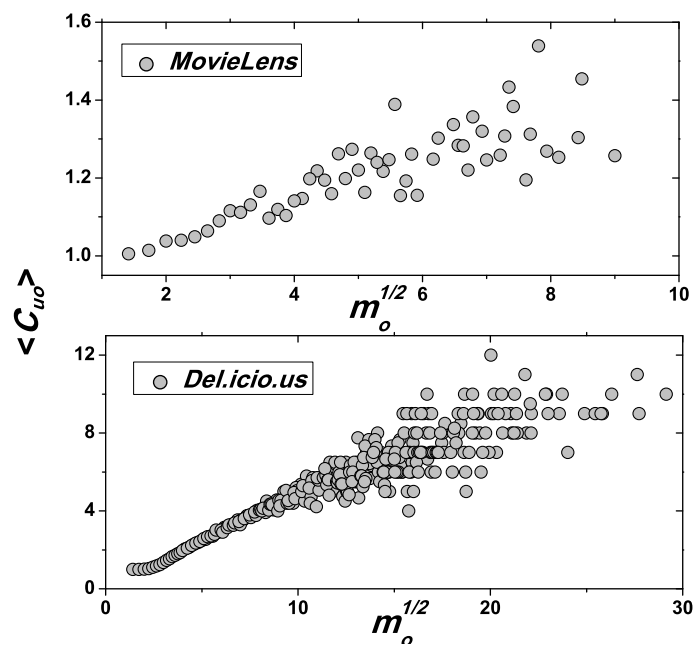}
\caption{ $\langle C \rangle$ as the function of $m^{1/2}$ for the two observed datasets, showing that the common feature, $\langle C \rangle$, is positively correlated with the object mass.} \label{fig:cm}
\end{center}
\end{figure}

\begin{figure}[ht]
\begin{center}
\includegraphics[width=13cm]{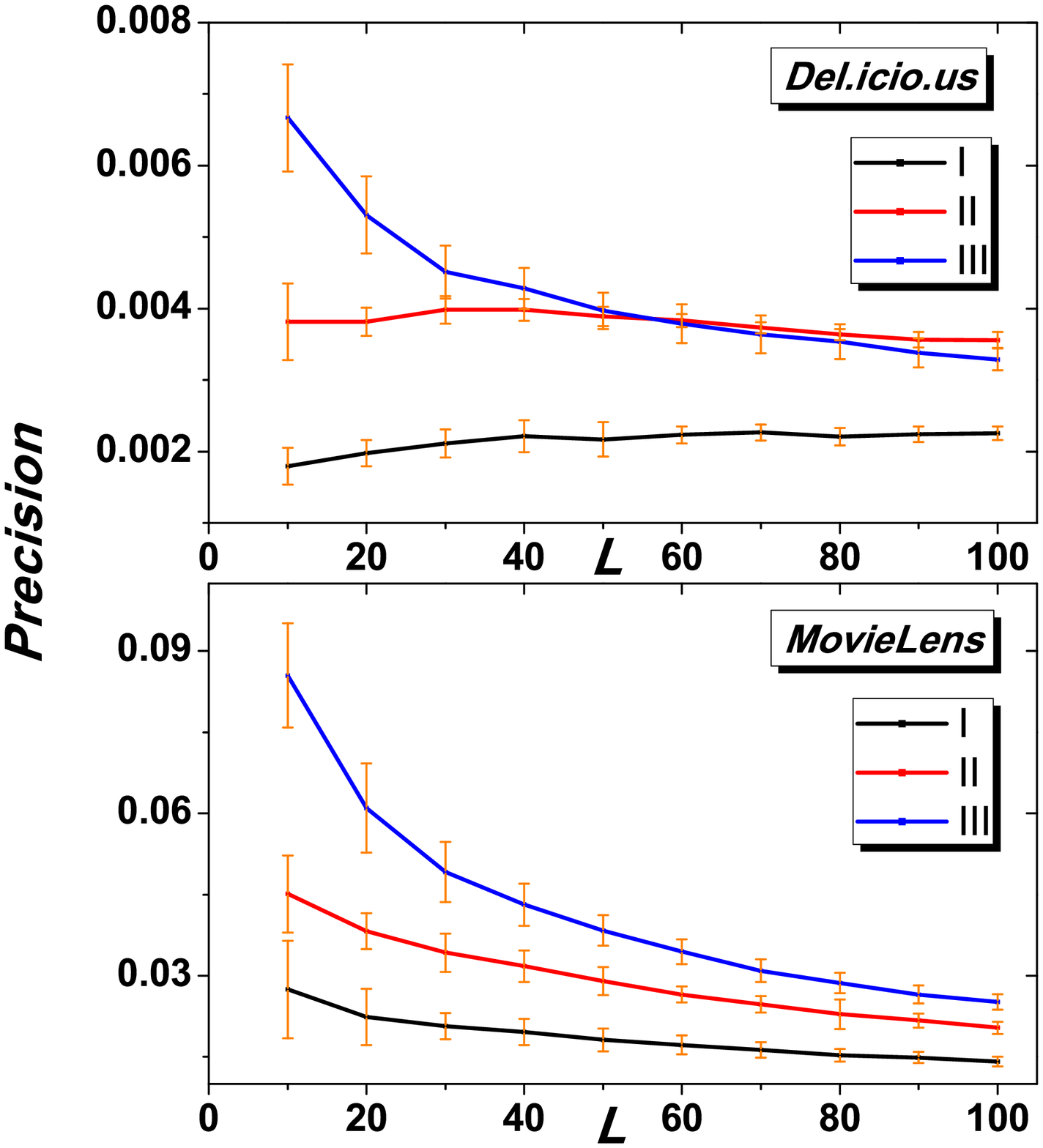}
\caption{(Color online) \emph{Precision} vs. recommendation length $L$ of the three algorithms for \emph{Del.icio.us} and \emph{Movielens}. The result is obtained by averaging over 50 independent realizations of random data division$N_c$, and yellow lines represent the error intervals. The parameter $\alpha$ for algorithm (III) is set to 0.001. Results on both datasets show that the gravity-model based algorithm (black) outperforms other two baselines.}\label{Fig2}
\end{center}
\end{figure}

\begin{figure}[ht]
\begin{center}
\includegraphics[width=13cm]{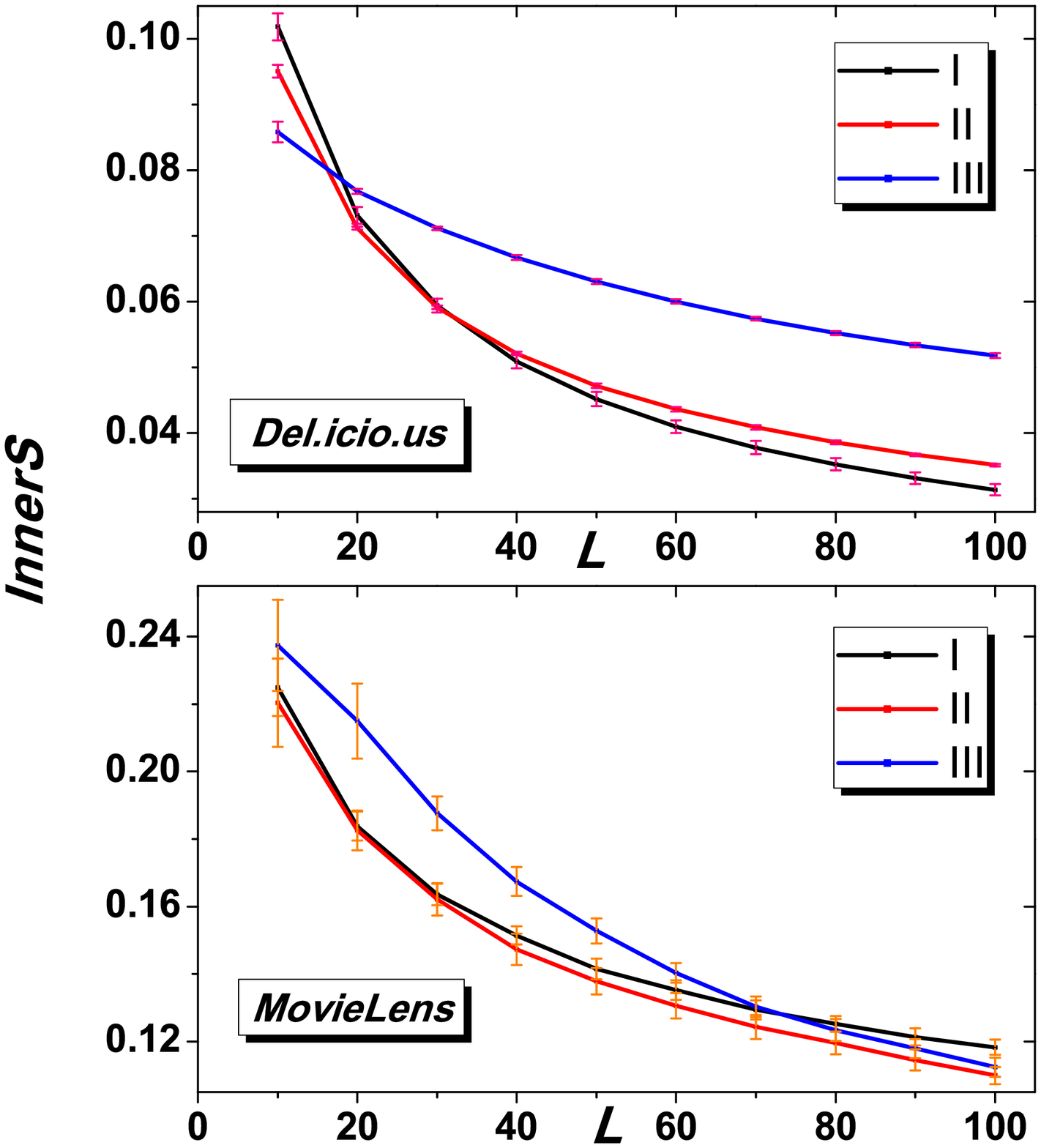}
\caption{(Color online) \emph{InnerS} vs. recommendation length $L$ of the three algorithms for \emph{Del.icio.us} and \emph{Movielens}. The result is obtained by averaging over 50 independent realizations of random data division, and yellow lines represent the error intervals. The parameter $\alpha$ for algorithm (III) is set to 0.001. Results on both datasets show that the gravity-model based algorithm (black) outperforms other two baselines.}\label{Fig4}
\end{center}
\end{figure}

Fig. \ref{Fig1} - Fig. \ref{Fig4} show the experimental results of those three metrics. In Fig. \ref{Fig1}, it can be seen that \emph{AUC} will decrease as $\alpha$ increases for both \emph{MovieLens} and \emph{Del.icio.us}. In addition, there are two stationary states of \emph{AUC} for large or small $\alpha$, which respectively correspond to the best and worst \emph{AUC}. In fact, Eq. (\ref{eq:GR}) can be transformed to,

\begin{equation}
\begin{array}{rcl}
 f^{(III)}_{o_j}&=&(C_{u_io_j}\sqrt{m_{u_i}m_{o_j}}+\alpha\sqrt{m_{u_i}m_{o_j}})^2 \\
\end{array}\label{eq:GR2}
\end{equation}

Note that, for a real dataset, $C_{u_io_j}$, $m_{u_i}$ and $m_{o_j}$ are finite. That is to say, for the extreme cases of Eq. (\ref{eq:GR2}): (a) $\alpha \gg C_{u_io_j}$ (but a finite value), $f_{o_j}^{(III)}$ is purely determined by $\sqrt{m_{u_i}m_{o_j}}$; (b) $\alpha \rightarrow \infty $, $f_{o_j}^{(III)} \rightarrow \infty$. $f_{o_j}^{(III)}$ will be the same for all objects, hence resulting in a random recommendation process. In the experiments, for the simplicity of calculation, we set $f_{o_j}^{(III)}$ as a constant; (c) $\alpha \rightarrow 0$, $f_{o_j}^{(III)}$ is hybridly determined by $C_{u_io_j}$ and $\sqrt{m_{u_i}m_{o_j}}$. In addition, as for a given recommendation process, once the target user $U_i$ is fixed, the value of $m_{u_i}$ will be the same for all examined objects. Consequently, the competition of $C_{u_io_j}$ and $m^{1/2}$ would finally determine whether object $O_j$ will be highly ranked and eventually recommended. For large $\alpha \gg C_{u_io_j}$, algorithm (III) degenerates to the object popularity priority first algorithm (so-called \emph{GRM} in \cite{zhou2007bipartite}). For $\alpha \rightarrow \infty$, algorithm (III) degenerates to random recommendation. For small $\alpha\rightarrow 0$, the final result is determined by the resultant force of $C_{u_io_j}$ and $m^{1/2}$, which we subsequently investigate in Fig. \ref{fig:cm}. It shows a clearly positive relationship between them. It means that the \emph{heavier} (corresponds to large $m_o$) an object is, the more chance it will be attracted by users with more common interests (corresponds to large $C_{uo}$), hence is more likely to be recommended by the proposed algorithm. In addition, Table \ref{tab:auc} shows the pure \emph{AUC} values of mass ($m_o$), common interest ($C_{uo}$) and algorithm (III). Indeed, it shows that $C_{uo}$ can significantly enhance the recommendation accuracy comparing with that based on pure object popularity. Furthermore, with incorporating the object popularity, the gravity law based method can achieve even better performance.


Fig. \ref{Fig2} and Fig. \ref{Fig4} show the results of \emph{Precision} and \emph{InnerS} as the function of the length of recommendation list, respectively. In both figures, algorithm (III) performs better than the other two baselines, especially for small $L$. Note that, in real applications, the number of recommended objects pushed to users could be very small (normally $L\in[1, 5]$ in real applications) due to the page limitation, the proposed algorithm might be very promising and useful in online applications.


\section*{Gravity Based Evolving Model}
In this section, we propose an evolving model to better understand the gravity effect on networks. Among various mechanisms driving the corresponding emergent properties, preferential attachment (PA)\cite{Barabasi199901} , which considers \emph{rich-get-richer}, is one of the most attractive models. However, the PA model only takes into account the mass of target nodes, while neglecting the underlying relationship between the two considering nodes. Consequently, we coherently present the gravity model to unify both node mass and common features, and compare it with two baseline models, \emph{ER} model \cite{ErdosP1960} and \emph{PA} model.

\subsection*{Model Description}

Using the divided training set extracted in the previous recommendation experiment, we build a static network (\emph{ST}) as the baseline for comparison. In the \emph{ST} network, nodes represent users, and one link will be created if the corresponding two users have collected at least one common object. Besides, each user has a weighted tag attribute vector. The final initialized network contains 648 vertices, 20,956 edges, and 1,382 tags. Comparatively, the other three observed evolving mechanisms at each step as following:

(I) \emph{ER} model. Select two nodes randomly and connect them if there is no link between them;

(II) \emph{PA} model. Select two nodes and connect them if there is no link between them. Each node $i$ is chosen according to its own degree $\frac{k_i}{\sum{k_i}}$. Initially, the degree of each node is set to one;

(III) \emph{GR} model. Select one node randomly, and link it to another node that is chosen based on the probability defined by Eq. (\ref{eq:GR}).

\subsection*{Results \& Analysis}

To give a solid and comprehensive evaluation of the \emph{GR} model, we employ five different metrics to characterize the properties of the resulting networks. The observed properties include (a) size of giant component \cite{ZhoutEPJB2009}; (b) assortativity \cite{Newman2002}; (c) clustering coefficient \cite{Watts199801}; (d) average distance \cite{Bouttier2003}; (e) degree  heterogeneity \cite{Albert2002}.

\begin{figure}[ht]
\begin{center}
\includegraphics[width=12cm]{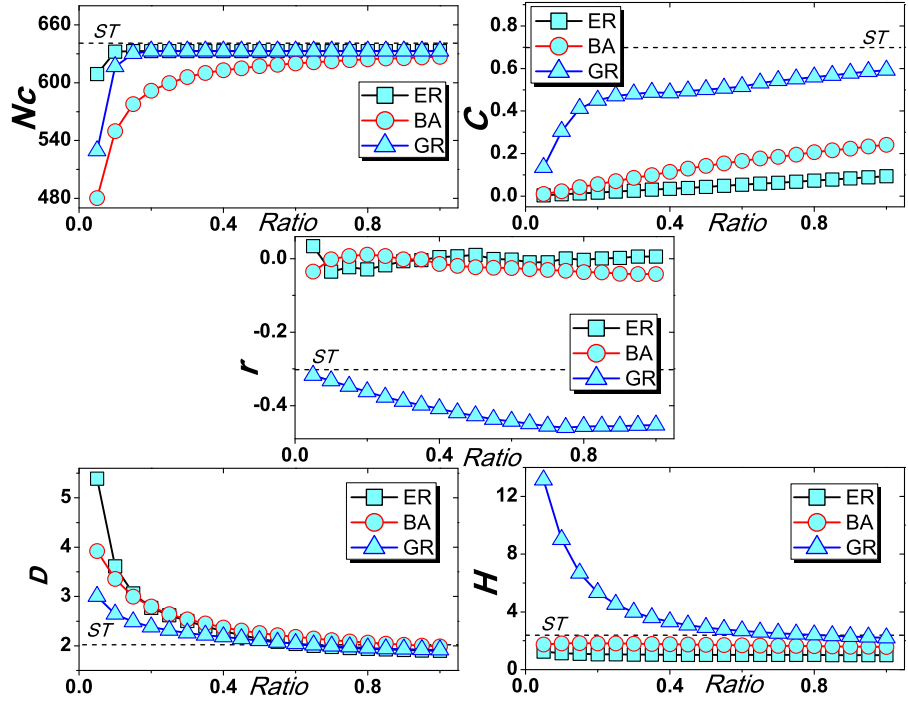}
\caption{(Color online) $N_c$, $C$, \emph{E}, \emph{r}, \emph{D} and \emph{H} as the function of ratio of added links. The result is obtained by averaging over 50 interdependent network realizations. The dash line highlights the corresponding result of \emph{ST} network. Results from five representative metrics show that the \emph{GR} model (blue triangle) is the best one to approach the original \emph{ST} network.}\label{Fig5}
\end{center}
\end{figure}

Table \ref{tab:model} shows the evolutionary results of corresponding models. In general, as \emph{ST} network is extracted from the real user-object bipartite network, it would naturally keep the original relationship of users' common interests. Therefore, it could be used as the baseline to compare the proposed models. That is to say, the model can better characterize the real-world evolutionary dynamics if its resulting properties are more similar with \emph{ST}. In Table \ref{tab:model}, it apparently shows that \emph{GR} performs much better than both \emph{ER} and \emph{PA} for all observed properties. Since \emph{GR} also considers users' interests by taking into account their preferences on assigning common tags, it would have a high probability to generate a more connected (corresponds to large $N_c$) and more clossness (corresponds to large $C$) network. Comparatively, the diverse topics (e.g. tags about different subjects) would make closer connections within the same community of similar topics, however, simultaneously increase the distance between nodes affiliated with different communities of different subjects, hence result in a larger network distance. In addition, the high degree heterogeneity ($H$) indicates that the network tends to be more disassortative (corresponds to negative $r$) \cite{ZhouS2007}. Furthermore, the dynamic link adding process (see Fig.~\ref{Fig5}) also shows the advantages of \emph{GR} in the five corresponding properties. In a word, \emph{GR} evolving mechanism indeed can result in a more real network with large clustering coefficient and network distance, high degree heterogeneity and a strongly disassortative linking pattern.

\section*{Conclusions and Discussion}
In this paper, we applied the classical gravity model in designing a new recommendation algorithm, considering the effects of both masses and common interests of two observed nodes. Experimental results on two real-world networks, \emph{MovieLens} and \emph{Del.ico.us}, demonstrated
that the proposed algorithm outperformed the previous two baseline algorithms. Furthermore, we adopted the gravity principle to build an evolving network to understand its advantage in information recommendation. Numerical analyses of five corresponding network properties proved the gravity mechanism can characterize the structure of real networks better than two baseline stochastic methods, \emph{ER} and \emph{PA} models. Therefore, the gravity-based algorithm can naturally provide more suitable results that can be recommended to appropriate users.

In brief, this work innovatively applies the gravity principle in information filtering and network evolving of online system. The results provide preliminary evidence of gravity effect on directly mining the hidden interests of users, and picking up relevant information. However, the underlying mechanism of the gravity effect on network-based algorithms and models still need further exploration.

\section*{Acknowledgments}
We thank Kun Chin Hu, Tao Zhou and Chengzhi Zhang for their valuable discussions.

\bibliography{refs}

\newpage

\begin{table}
\centering \caption{Basic statistical properties of the two datasets. $|U|$, $|O|$, $|T|$ and $D$ respectively represent the number of users, objects, tags and tagging actions, and $S=\frac{D}{|U|\times|O|}$ denotes the data sparsity.}
\setlength\tabcolsep{18pt}
\begin{tabular}{ccccccc}
\hline
\hline
Data & $|U|$ & $|O|$ & $|T|$ & $D$ & $S$ \\
\hline
\emph{Mov.} & 648 & 1,590 & 1,382 & 9,749 & 9.5 $\times10^{-3}$   \\
\hline
\emph{Del.} & 4,902 & 36,224 & 10,584 & 192,487 & 1.1 $\times10^{-3}$ \\
\hline
\hline
\end{tabular}\label{tab:data}
\end{table}
\begin{table}
\centering \caption{Comparisons of \emph{AUC} results of respectively considering the effects of mass ($m_o$), common interest ($C_uo$), and as well as three algorithms (algorithm I, II and III). The result is obtained by averaging over 50 independent realizations of random data division, and the three digital numbers behind the signs are the corresponding error intervals. The parameter $\alpha$ for algorithm (III) is set to 0.001.}
\setlength\tabcolsep{10pt}
\begin{tabular}{cccccc}
\hline
\hline
Data & $m_o$ & $C_{uo}$ & (I)& (II)& (III) \\
\hline
\emph{Mov.} & $0.706\pm 0.030$ & $0.882\pm 0.017$ & $0.870\pm0.010$ & $0.877\pm 0.016$ &  $\mathbf{0.889\pm0.009}$ \\
\hline
\emph{Del.} & $0.612\pm 0.004$ & $0.744\pm 0.002$ & $0.791\pm 0.005$ & $0.804\pm 0.003$  & $\mathbf{0.797\pm 0.004}$ \\
\hline
\hline
\end{tabular} \label{tab:auc}
\end{table}

\newpage
\begin{table}
\centering \caption{Evolutionary results of four corresponding networks. $N_c$ represents the size of the giant component, $C$ denotes the clustering coefficient, $r$ and $D$ are respectively the assortative coefficient and average distance of network, and $H=\frac{<k^2>}{<k>^2}$ denotes the network heterogeneity. In the last three rows, it presents both real value of corresponding metric and the error interval (separated by slash), which is calculated as: $\frac{|V_{real}-V_{ST}|}{V_{ST}} \times 100\%$, where $V_{real}$ is the metric value of current model and $V_{ST}$ is the corresponding value of \emph{ST} network. Each value is obtained by averaging over 50 interdependent network realizations.}
\setlength\tabcolsep{12pt}
\begin{tabular}{cccccc}
\hline
\hline
 & $N_c$ & $C$ & \emph{r} & \emph{D} & \emph{H} \\
\hline
\emph{ST} & 641 & 0.698 &  -0.302 & 2.021 & 2.390 \\
\hline
\emph{ER} & 633/1.25\% & 0.094/86.5\% & 0.006/102.0\% & 1.898/6.1\% & 1.009/57.8\% \\
\hline
\emph{PA} & 627/21.8\% & 0.241/65.5\% & -0.042/86.1\% & \textbf{1.988/1.6\%} & 1.570/33.4\% \\
\hline
\emph{GR} & \textbf{633/1.25\%} & \textbf{0.592/15.2\%} & \textbf{-0.453/50.0\%} & 1.913/5.3\% & \textbf{2.201/7.6\%} \\
\hline
\hline
\end{tabular} \label{tab:model}
\end{table}

\clearpage

\section*{Figure Legends}

%
%

\end{document}